\documentclass[aps,prb,reprint,amsmath,amssymb,superscriptaddress,floatfix]{revtex4-2}
\usepackage{graphicx}
\usepackage{bm}
\usepackage{natbib}
\usepackage{url}
\usepackage{textcase}
\usepackage{amssymb}
\usepackage{amsmath}
\usepackage{physics}
\usepackage{appendix}
\usepackage{xcolor}
\usepackage{comment}
\usepackage[colorlinks=true,linkcolor=blue,urlcolor=blue,citecolor=blue,anchorcolor=blue]{hyperref}
\usepackage[normalem]{ulem}

\newcommand\comeout[1]\null

\begin{document}

\title{
Generating collective spin cat states via photon-number measurements near the Dicke critical point
}

\author{Taiga Nakamoto}
\email{taiganakamoto@g.ecc.u-tokyo.ac.jp}
\affiliation{Department of Physics, University of Tokyo, 7-3-1 Hongo, Tokyo 113-0033, Japan}

\author{Shohei Imai}
\affiliation{Department of Physics, University of Tokyo, 7-3-1 Hongo, Tokyo 113-0033, Japan}

\author{Kazuaki Takasan}
\affiliation{Department of Applied Physics, University of Tokyo, 7-3-1 Hongo, Tokyo 113-8656, Japan}

\date{\today}

\begin{abstract}
We propose a method for generating collective spin cat states in a cavity-coupled atomic ensemble by exploiting strong light–matter entanglement and anti-squeezing associated with the superradiant phase transition. 
We numerically and analytically demonstrate that the cat states can be heralded by photon-number measurement on the ground state of the Dicke model. 
The near-critical regime enhances both the cat-state size and the probability of obtaining larger photon-number outcomes, and outcomes with larger photon numbers yield even larger cat states. 
We also show that a thermodynamic-limit analysis clarifies the generation mechanism and connects it to a natural light–matter analogue of generalized photon subtraction for optical cat-state generation. 
These results suggest that exploiting criticality in strongly coupled light–matter systems could open new directions for matter-based many-body quantum technologies.
\end{abstract}

\maketitle

\section{Introduction}
A cat state is a non-classical state formed by a quantum superposition of two macroscopically distinct states~\cite{schrodinger_Gegenwartige_1935,deleglise_Reconstruction_2008}, such as bosonic coherent states of light or coherent spin states of atomic ensembles.
They have attracted considerable interest because of their potential applications in quantum-enhanced metrology, quantum information processing, and quantum computation~\cite{munro_Weakforce_2002,joo_Quantum_2011,huang_Quantum_2015,pezze_Quantum_2018,ralph_Quantum_2003,ofek_Extending_2016}.
For these applications, the quality of a cat state is governed by two complementary aspects: the distinguishability of the two superposed components and the preservation of coherence between them~\cite{cochrane_Macroscopically_1999,frowis_Macroscopic_2018,lescanne_Exponential_2020}.
The separation between the components sets the cat-state size, while the remaining coherence determines the fidelity to an ideal quantum superposition. 
It is therefore important to develop methods for generating cat states that are simultaneously large and high-fidelity across a wide range of physical platforms.

In optical systems, high-fidelity generation of such states is typically based on heralding schemes~\cite{dakna_Generating_1997,ourjoumtsev_Generating_2006,neergaard_Generation_2006,takahashi_Generation_2008,gerrits_Generation_2010,huang_Optical_2015,sychev_Enlargement_2017}.
In these schemes, two squeezed states are first generated by nonlinear optical processes and then mixed at a beam splitter.
A photon-number measurement on one output mode then probabilistically heralds the preparation of a non-classical cat state in the other output mode.
Non-classicality of the cat state is extracted through the measurement process in such a protocol.
The underlying idea of such a heralding scheme is not limited to optics. 
In general, measurements on one subsystem can induce non-classical states in the other entangled subsystem.

In particular, hybrid systems of atoms and light are promising platforms for measurement-induced non-classical states of atomic ensembles~\cite{saito_Measurementinduced_2003,sørensen_Measurement_2003}. 
It has been demonstrated that, after entangling an atomic ensemble with light, photon-number measurement on the optical field induces a non-classical state in the atomic ensemble \cite{mcconnell_Entanglement_2015}.
This approach offers a potentially powerful strategy for generating a collective spin cat state (SCS)~\cite{massar_Generating_2003,genes_Generating_2006,mcconnell_Generating_2013,pettersson_Lightmediated_2017,davis_Painting_2018,imai_Macroscopic_2026}, which is a superposition of two distinct coherent spin states.
For such a strategy to work effectively, it is crucial to enhance both the generation probability and the resulting cat-state size, which quantifies the distance between the two coherent spin states.
To achieve these enhancements, strong light-matter entanglement and anti-squeezing play an important role.

\begin{figure}[t]
  \includegraphics[width=0.48\textwidth]{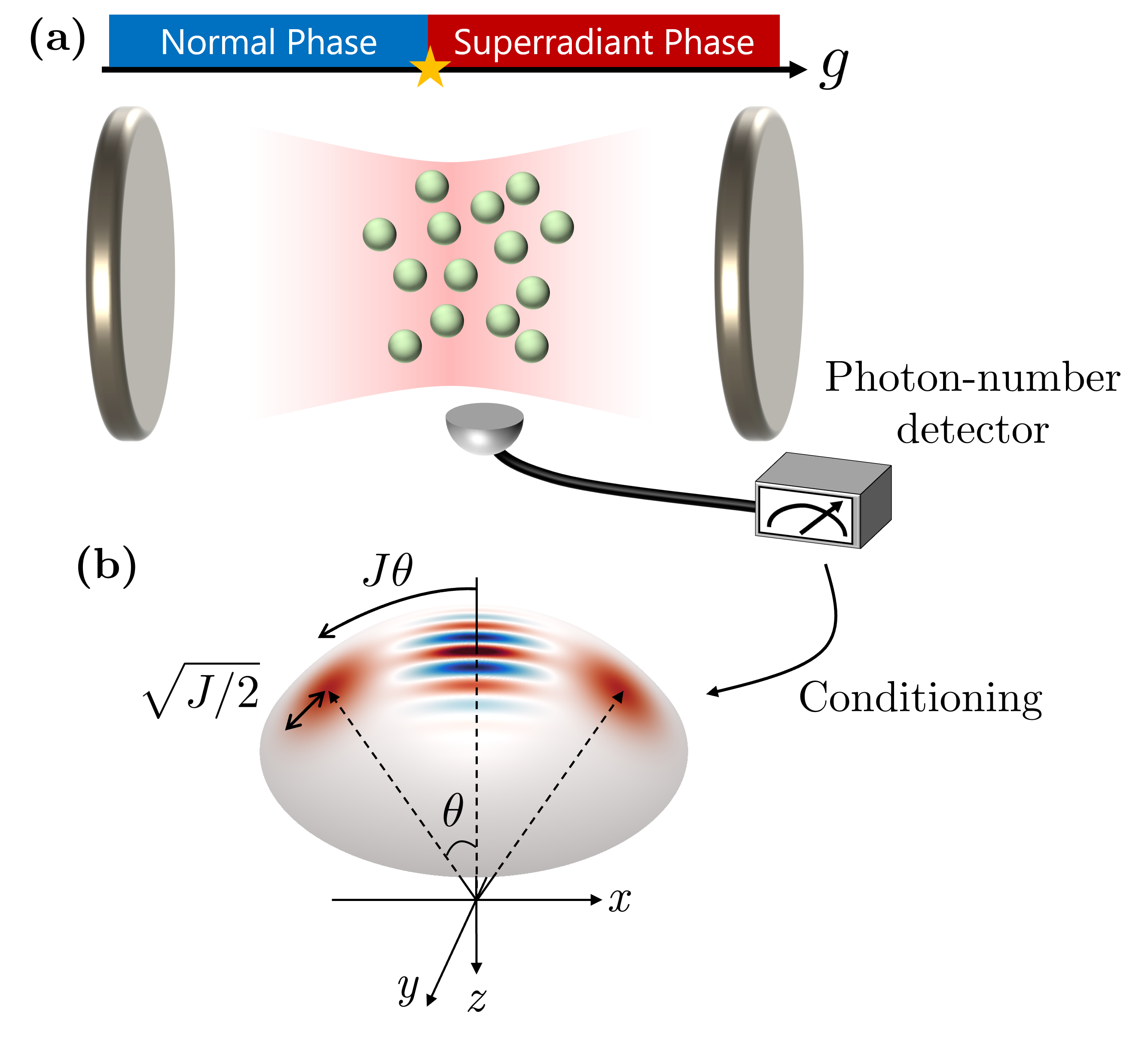}
  \caption{
    (a) Schematic picture of the protocol for generating SCSs.
    We first prepare an atomic ensemble that is strongly coupled to the cavity field near the critical point of the superradiant phase transition.
    We then perform a photon-number measurement on the cavity field; conditioning on a photon-number outcome heralds the preparation of an SCS in the atomic ensemble.
    (b) Schematic illustration of the Wigner function of the SCSs.
    We focus on the SCS, which is a superposition of two coherent spin states pointing in the directions ($+\theta,0$) and ($-\theta,0$).
    We characterize the size of the SCS as an arc length $J\theta$ normalized by the quantum fluctuation $\sqrt{J/2}$ of the coherent spin states.
  } \label{fig:concept}
\end{figure}

In this paper, we focus on the superradiant phase transition~\cite{hepp_Equilibrium_1973,wang_Phase_1973}, which exhibits diverging light-matter entanglement entropy and large anti-squeezing~\cite{lambert_Entanglement_2004}.
We present a protocol for producing SCSs by measuring the number of photons in the cavity field near the critical point (Fig.~\hyperref[fig:concept]{1(a)}).
Approaching the critical point enhances the size of the SCS and the measurement probability of a larger photon number.
Measurement of larger photon number further amplifies the size of the resulting SCS.
Moreover, we show that the mechanism for SCS generation is closely related to generalized photon subtraction in quantum optics~\cite{takase_Generation_2021}, a heralded scheme proposed for high-rate generation of optical cat states.
These results suggest that the criticality associated with the superradiant phase transition can be exploited as a resource for measurement-induced non-classicality. 
By combining this critical enhancement with photon-number measurement, our protocol offers a possible route to well-resolved and high-fidelity SCSs in atomic ensembles, which are promising quantum states for quantum-enhanced metrology and quantum information processing~\cite{munro_Weakforce_2002,joo_Quantum_2011,huang_Quantum_2015,pezze_Quantum_2018,ralph_Quantum_2003,ofek_Extending_2016}.

This paper is organized as follows.
In Sec.~\ref{sec:model_and_method}, we introduce the model Hamiltonian and the method for characterizing the SCS.
In Sec.~\ref{sec:proposal}, we propose a protocol for generating SCSs by photon-number measurement near the critical point of the superradiant phase transition.
In Sec.~\ref{sec:size_and_probability}, we quantify the size of the SCS and the measurement probability of photons by numerical simulations for finite size systems.
In Sec.~\ref{sec:thermodynamic_limit}, we provide an analytical understanding of the mechanism for SCS generation based on an analysis in the thermodynamic limit.
Finally, we discuss the implications of our results and conclude in Sec.~\ref{sec:discussion_conclusion}.

\section{Model and Methods} \label{sec:model_and_method}

\begin{figure*}[t!]
  \includegraphics[width=0.95\textwidth]{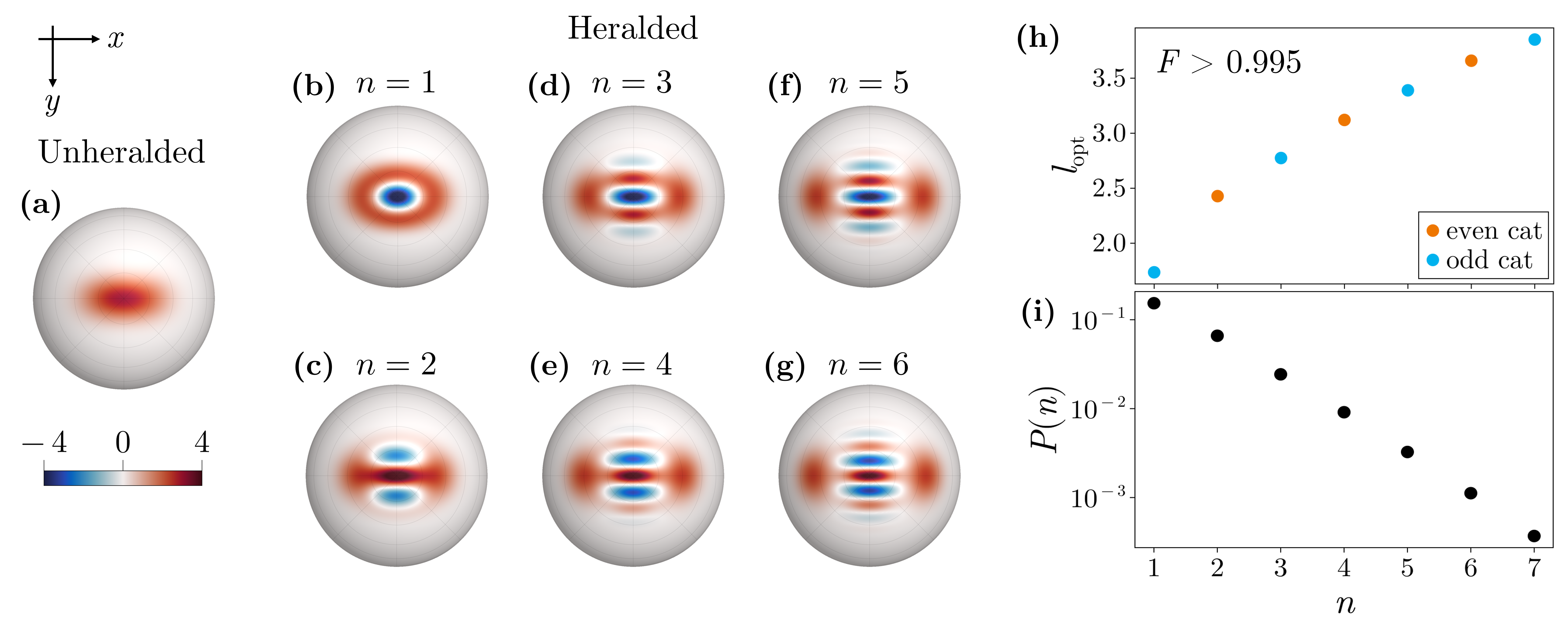}
  \caption{
    The Wigner function $W(\theta,\phi)$ around $\theta=0$ of (a) the unheralded state $\rho = \Tr_\text{photon}[\ket{G}\bra{G}]$ and (b)-(g) the $n$-photon heralded atomic state $\ket{\psi_n}$ for the outcomes $n=1, \ldots, 6$.
    The Wigner function of $\rho$ is Gaussian-like, while that of $\ket{\psi_n}$ has negative values and interference patterns.
    The interference patterns in the Wigner function indicate the production of SCSs.
    Heralding converts the Gaussian-like state into a non-Gaussian state with parity-dependent interference fringes.
    (h) The size of the SCS $l_\text{opt}$ as a function of $n$.
    When $n$ is even (odd), the parity of the SCS is even (odd).
    $l_\text{opt}$ increases as the measured photon number $n$ increases.
    The fidelities $F(\theta_\text{opt})$ between the ideal SCS and the $n$-photon heralded state are more than 0.995 for all $n$.
    (i) The measurement probability $P(n)$ as a function of $n$.
    As $n$ increases, the probability decreases, which indicates a trade-off between the size of the SCS and the measurement probability.
    Here, we set $N=30$ and $g=g_c$.
  } \label{fig:n_photon_dependence}
\end{figure*}

\subsection{Dicke model and superradiant phase transition} \label{subsec:Dicke_model}
We consider the Dicke model, which describes an ensemble of $N$ two-level atoms interacting with a single-mode cavity field.
The two-level atoms are represented as spin-1/2 particles, and the collective spin operators are defined as $\hat{J}_\alpha = \sum_{i=1}^N \hat{\sigma}_\alpha^{(i)}/2$ for $\alpha = x,y,z$, where $\hat{\sigma}_\alpha^{(i)}$ is the Pauli operator for the $i$-th atom.
The Hamiltonian of the Dicke model is given by
\begin{align}
  \hat{H} \!=\! \omega_\text{cav}\hat{a}^\dagger \hat{a} + \omega_\text{atom} \left(\!\hat{J}_z+\frac{N}{2}\right) \!+\! \frac{g}{\sqrt{N}} (\hat{a}^\dagger + \hat{a})(\hat{J}_+ \!+\! \hat{J}_-), \label{eq:Dicke_Hamiltonian}
\end{align}
where $\hat{a}^\dagger$ and $\hat{a}$ are the creation and annihilation operators of the cavity field with resonance frequency $\omega_\text{cav}$, respectively.
$\hat{J}_\pm=\hat{J}_x \pm i\hat{J}_y$ are the raising and lowering collective spin operators describing the two-level atoms with the transition frequency $\omega_\text{atom}$.
$g$ is the coupling strength between the atoms and the cavity field, and the factor of $1/\sqrt{N}$ ensures a well-defined thermodynamic limit.
In the main text, we consider the resonant case $\omega_\text{cav} = \omega_\text{atom} =: \omega$ to focus on the critical behavior of the system.
In Appendix~\ref{off_resonant}, we confirm that the mechanism for SCS generation also works in the off-resonant case $\omega_\text{cav} \neq \omega_\text{atom}$.

The states of the total system are expanded in the basis $\{ \ket{n} \otimes \ket{J,m} \}$, where $\ket{n}$ are number states of the cavity field and $\ket{J,m}$ are the Dicke states of the collective spin.
Since the Hamiltonian conserves the total spin $J=N/2$, the collective spin degrees of freedom can be described by an $(N+1)$-dimensional Hilbert space $\{ \ket{J,m} \mid m=-J,-J+1,\ldots,J \}$.
These states satisfy $\hat{J}^2 \ket{J,m} = J(J+1) \ket{J,m}$ and $\hat{J}_z \ket{J,m} = m \ket{J,m}$.
Moreover, the Hamiltonian conserves the excitation parity, $(-1)^{n+m+J}$, which separates the Hilbert space into two sectors with even and odd parity.
The ground state of the system belongs to the even parity sector before the superradiant phase transition~\cite{emary_Chaos_2003}.

The Dicke model exhibits the superradiant phase transition at a critical coupling strength $g_c = \sqrt{\omega_\text{cav} \omega_\text{atom}}/2$ in the thermodynamic limit $N \to \infty$~\cite{hepp_Equilibrium_1973,wang_Phase_1973}.
For $g < g_c$, the system is in the normal phase, where the cavity field has no macroscopic occupation and the atomic ensemble remains unexcited.
For $g > g_c$, the system enters the superradiant phase, where the cavity field acquires a macroscopic occupation and the atomic ensemble exhibits macroscopic excitation.
At the critical point $g = g_c$, the system exhibits critical behavior, including diverging light-matter entanglement entropy~\cite{lambert_Entanglement_2004} and a high degree of squeezing~\cite{emary_Chaos_2003,shapiro_Universal_2020,hayashida_Perfect_2023}.


\subsection{SCS and its characterization} \label{subsec:characterization_SCS}
SCSs are defined as superpositions of two coherent spin states pointing in different directions.
The coherent spin states are given by
\begin{align}
  \ket{\theta,\phi} = \hat{R}(\theta,\phi) \ket{J,-J},
\end{align}
where $\ket{J,-J} = \ket{\downarrow, \downarrow, \ldots, \downarrow}$ is the spin ground state, and $\hat{R}(\theta,\phi) = e^{-i\theta\sin\phi \hat{J}_x + i\theta\cos\phi \hat{J}_y}$ rotates the state $\ket{J,-J}$ into the direction $(\sin\theta \cos\phi, \sin\theta \sin\phi, -\cos\theta)$.
In this paper, we focus on the SCS of the following form:
\begin{align}
  \ket{\text{cat}(\theta)} \propto \ket{+\theta,0} \pm \ket{-\theta,0}, \label{eq:def_SCS}
\end{align}
where $\ket{+\theta,0}$ and $\ket{-\theta,0}$ are two coherent spin states pointing in the directions as shown in Fig.~\hyperref[fig:concept]{1(b)}.
$\pm$ corresponds to the even and odd SCS, respectively.

To quantify how large the SCS is, we first define the fidelity between a certain state $\ket{\psi}$ and the ideal SCS $\ket{\text{cat}(\theta)}$.
The fidelity $F(\theta) = |\bra{\text{cat}(\theta)}\ket{\psi}|^2$ quantifies the closeness between $\ket{\psi}$ and $\ket{\text{cat}(\theta)}$.
We compute the optimal angle $\theta_\text{opt}$ by maximizing the fidelity as
\begin{align}
  \theta_\text{opt} = \text{argmax}_{\theta} F(\theta). \label{eq:theta_opt}
\end{align}
$\theta_\text{opt}$ characterizes the size of the SCS that best matches $\ket{\psi}$.
Next, we consider the arc length $J\theta_\text{opt}$, which quantifies the displacement of each coherent spin state from the spin ground state $\ket{J,-J}$. 
Since the coherent spin states have a quantum fluctuation $\sqrt{J/2}$ perpendicular to the spin direction as illustrated in Fig.~\hyperref[fig:concept]{1(b)}, we normalize the arc length by the quantum fluctuation.
Thus, we focus on the following quantity as a measure of the size of the SCS:
\begin{align}
  l_\text{opt} = \frac{J\theta_\text{opt}}{\sqrt{J/2}} = \sqrt{N} \theta_\text{opt}. \label{eq:normalized_arc_length}
\end{align}
Due to the normalization, $l_\text{opt}$ remains finite even in the thermodynamic limit $N \to \infty$.
If $l_\text{opt} \gg 1$, the two components of the SCS are well separated, while if $l_\text{opt}<1$, there is a significant overlap between the two coherent spin states.

To visualize the non-classicality of a state, we also compute the spin Wigner function~\cite{brif_Phasespace_1999,davis_Wigner_2021}.
Given the density-matrix element $\rho_{mm'} = \bra{J,m}\hat{\rho}\ket{J,m'}$, the spin Wigner function is defined as
\begin{align}
  W(\theta,\phi) = \sqrt{\frac{2J+1}{4\pi}}\sum_{k=0}^{2J} \sum_{q=-k}^k t_{kq} Y_{kq}(\theta,\phi), \\
  t_{kq} = \sum_{m,m'=-J}^J \rho_{mm'} \sqrt{\frac{2k+1}{2J+1}} \braket{J,m';k,q}{J,m},
\end{align}
where $Y_{kq}(\theta,\phi)$ is the spherical harmonic, and $\braket{J,m';k,q}{J,m}$ is the Clebsch-Gordan coefficient.
The Wigner function is normalized as $\int^{\pi}_0 d\theta \int^{2\pi}_0 \sin\theta d\phi W(\theta,\phi)=1$.
The Wigner function of the SCSs of Eq.~\eqref{eq:def_SCS} has an interference pattern with negative values around $\theta=0$ as shown in Fig.~\hyperref[fig:concept]{1(b)}. 

\section{Photon-number heralding protocol} \label{sec:proposal}
We propose a method for generating SCSs exploiting the critical point of the superradiant phase transition as illustrated in Fig.~\hyperref[fig:concept]{1(a)}.
We first prepare the ground state of the Dicke model $\ket{G}$ near the critical point $g\lesssim g_c$.
We then perform a photon-number measurement on the cavity field to herald an atomic state.
Our main finding is that the photon-number heralded atomic state is very close to a SCS while the unheralded atomic state is not.

To demonstrate the cat-state generation method, we use the exact diagonalization with photon-number cutoff $n_\text{cutoff}=50$.
We check the convergence of the results with respect to the cutoff (see Appendix~\ref{sec:numerical_convergence} for the details).

We first show the results for the unheralded case.
The unheralded atomic state is described by the reduced density matrix after tracing out the photonic degrees of freedom $\hat{\rho} = \Tr_\text{photon}[\ket{G}\bra{G}]$.
Figure~\hyperref[fig:n_photon_dependence]{2(a)} shows the Wigner function of $\hat{\rho}$ for $N=30$ and $g=g_c$.
The Wigner function is anti-squeezed in the $x$-direction and slightly squeezed in the $y$-direction, which are precursors of the superradiant phase transition~\cite{emary_Chaos_2003}.
However, the Wigner function still has a Gaussian-like form.
The unheralded state does not exhibit Wigner negativity or cat-like interference patterns.

Next, we show the results for the heralded case.
The $n$-photon measurement projects the total state $\ket{G}$ onto the atomic state $\ket{\psi_n}$, which is given by 
\begin{align}
  \ket{\psi_n} = \frac{\bra{n}\ket{G}}{\|\bra{n}\ket{G}\|}. \label{eq:def_psi_n}
\end{align}
Figures~\hyperref[fig:n_photon_dependence]{2(b)-(g)} show the Wigner function of $\ket{\psi_n}$ for $n=1, \ldots, 6$, $N=30$, and $g=g_c$.
Two Gaussian peaks appear on the $xz$ plane, corresponding to the two coherent-spin components of the cat state.
The heralded states exhibit Wigner negativity around $\theta=0$ and interference patterns, which are hallmarks of the SCS. 
The photon-number heralding converts the unheralded Gaussian-like state into a non-Gaussian SCS.

We note that the parity of the measured photon number $n$ determines the parity of the resulting SCS.
The even (odd) parity SCS is a superposition of $\ket{J,m}$ with even (odd) $m+J$.
Since the parity of $(-1)^{n+m+J}$ is even for the ground state, the parity of $n$ is the same as that of $m+J$.
The even (odd) parity SCS has a positive (negative) value at $\theta=0$ as shown in Fig.~\hyperref[fig:n_photon_dependence]{2(b)-(g)}. 

In heralding schemes for cat-state generation, the size of the cat state and the success probability of the heralding are both important.
We quantify the size of the SCS and the measurement probability of photons by numerical simulations for finite size systems in Section~\ref{sec:size_and_probability}.
Then, we provide an analytical understanding of the mechanism for SCS generation based on an analysis in the thermodynamic limit in Section~\ref{sec:thermodynamic_limit}.

\section{Size of SCS and measurement probability} \label{sec:size_and_probability}

In this section, we investigate the size of the SCS and the measurement probability of photon number by numerical simulations for finite size systems.
As indicated in Fig.~\hyperref[fig:n_photon_dependence]{2(b)-(g)}, the size of the SCS increases as the measured photon number increases.
We quantify the size of the SCS by computing $l_\text{opt}$ defined in Eq.~\eqref{eq:normalized_arc_length}.
We compare $\ket{\psi_n}$ for even (odd) $n$ with the even (odd) SCS $\ket{\text{cat}(\theta)}$ to compute $l_\text{opt}$.
Figure~\hyperref[fig:n_photon_dependence]{2(h)} shows the $n$-dependence of $l_\text{opt}$.
We find that the size of the SCS $l_\text{opt}$ increases with $n$.
We check that the fidelities between $\ket{\psi_n}$ and $\ket{\text{cat}(\theta_\text{opt})}$ are more than 0.995.
Therefore, the states $\ket{\psi_n}$ are close to the ideal SCSs, and the size of the SCS increases as the measured photon number $n$ increases. 

On the other hand, it is less likely to generate a larger SCS by measuring a larger number of photons.
Figure~\hyperref[fig:n_photon_dependence]{2(i)} shows the $n$-dependence of the measurement probability $P(n)=\|\bra{n}\ket{G}\|^2$.
The data show that the probability $P(n)$ exponentially decreases as $n$ increases.
There is a trade-off between the size of the SCS and the measurement probability.

\begin{figure}[t!]
  \includegraphics[width=0.48\textwidth]{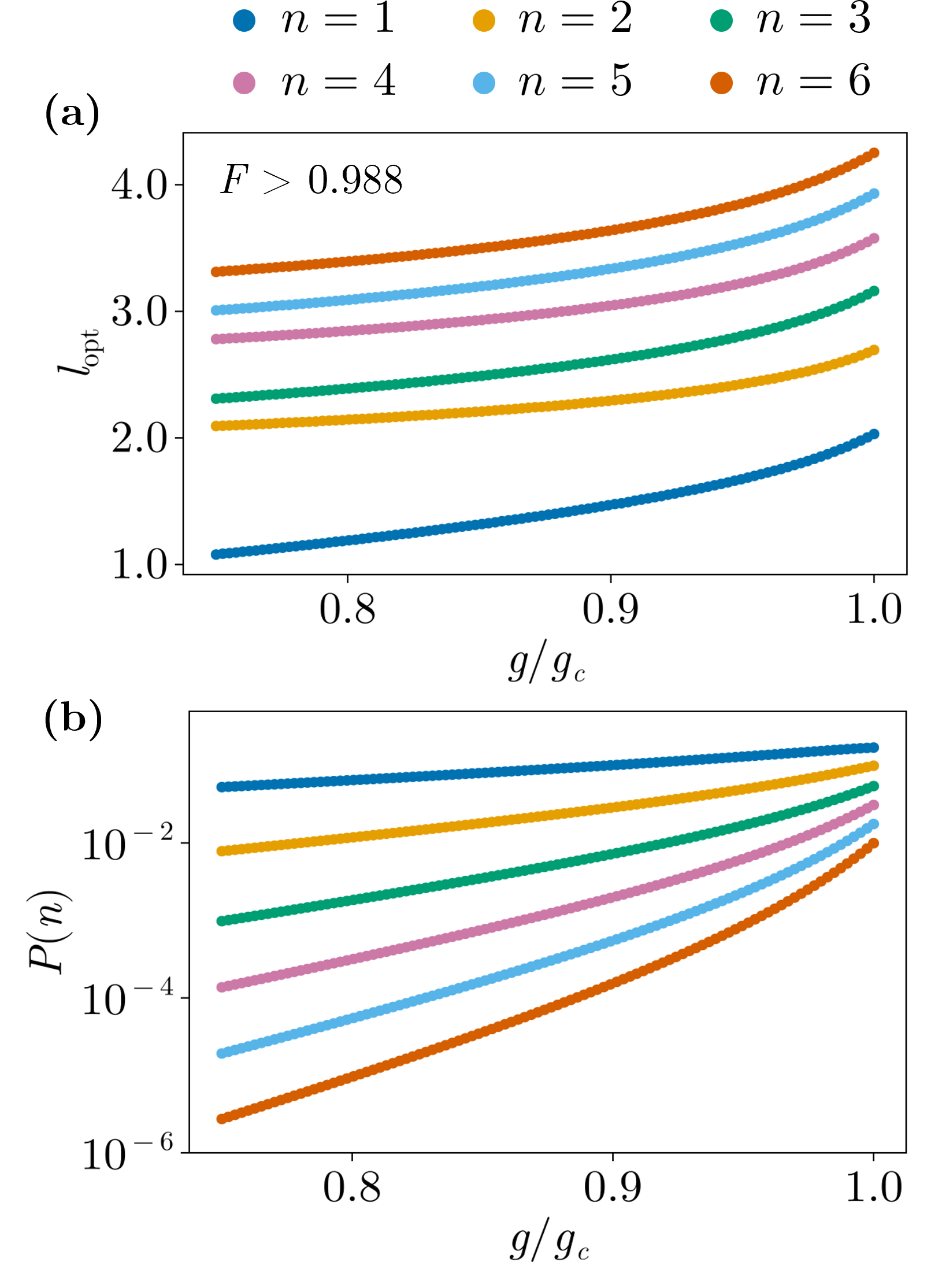}
  \caption{
    (a) The size of the SCS $l_\text{opt}$ for $n=1, \dots, 6$ as a function of the coupling strength $g$.
    As $g$ approaches the critical point $g_c$, $l_\text{opt}$ increases.
    Especially, $l_\text{opt}$ is significantly enhanced near the critical point $g=g_c$.
    The fidelities $F(\theta_\text{opt})$ are more than 0.988 for all the data points.
    (b) The measurement probability of $n$ photons $P(n)$ as a function of $g$.
    As $g$ approaches $g_c$, $P(n)$ exponentially increases.
    Here, we set $N=200$.
  } \label{fig:g_dependence}
\end{figure}

Next, we investigate the dependence of $\ket{\psi_n}$ on the coupling strength $g$.
Figure~\hyperref[fig:g_dependence]{3(a)} shows the $g$-dependence of $l_\text{opt}$ for $n=1, \dots, 6$ and $N=200$. 
We find that $l_\text{opt}$ grows as $g$ approaches the critical point $g_c$ for all $n$.
It is observed that $l_\text{opt}$ significantly increases near the critical point $g=g_c$.
Therefore, the criticality associated with the superradiant phase transition can serve as a resource for producing large SCSs.

The measurement probability is also substantially enhanced near the critical point.
Figure~\hyperref[fig:g_dependence]{3(b)} shows the $g$-dependence of $P(n)$ for $n=1, \dots, 6$ and $N=200$.
We find that $P(n)$ increases exponentially as $g$ approaches $g_c$ for all $n$.
In particular, increasing $g$ is more effective for measuring a larger $n$.
The probability for $n=2, \dots, 6$ photons become closer to that of $n=1$ photon as the system approaches the critical point.
As already mentioned in Fig.~\hyperref[fig:n_photon_dependence]{2(h)}, $l_\text{opt}$ significantly increases as $n$ increases.
Therefore, the enhancement of the measurement probability near the critical point also helps to generate a larger SCS.

\begin{figure}[t!]
  \includegraphics[width=0.48\textwidth]{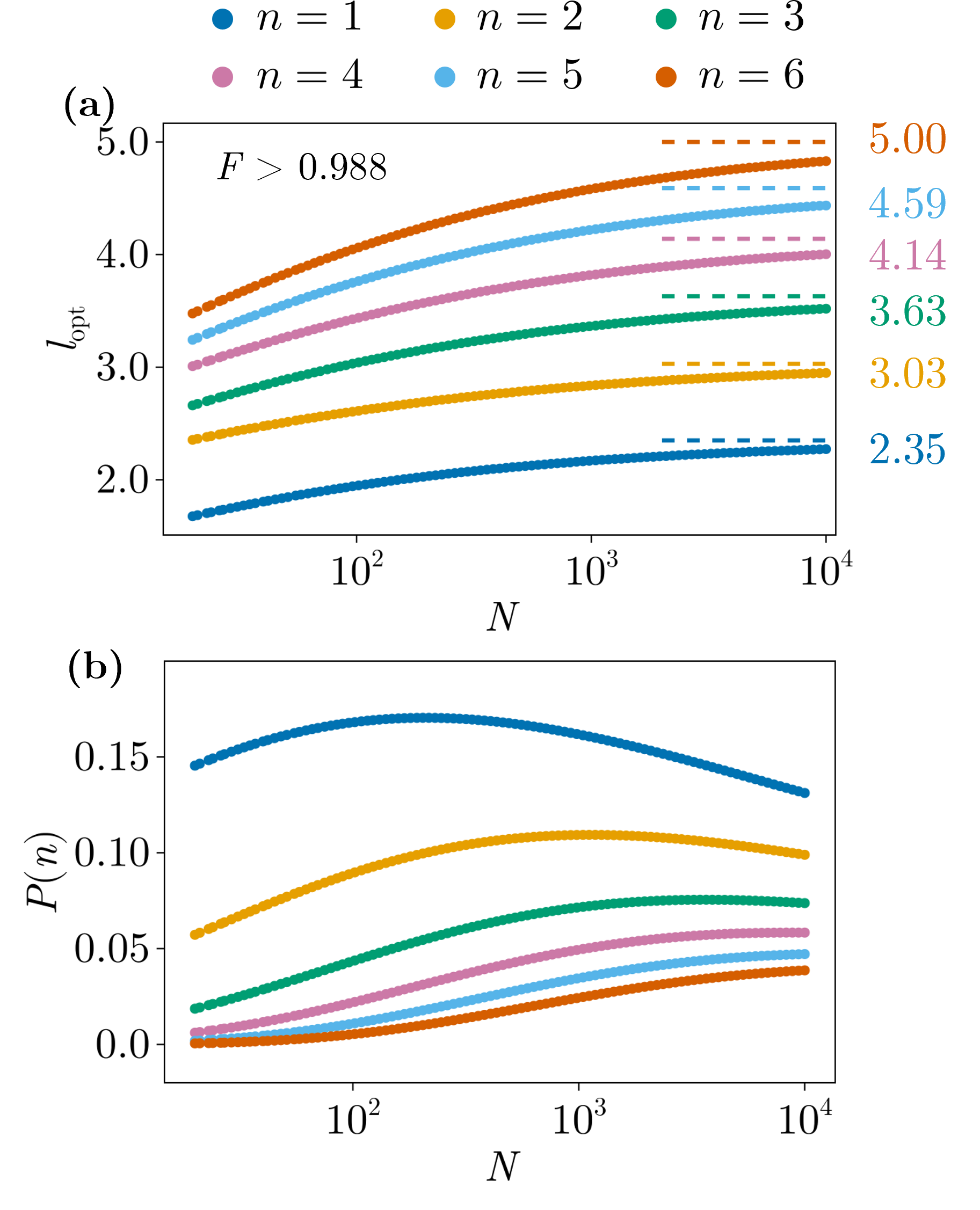}
  \caption{
    (a) The size of the SCS $l_\text{opt}$ for $n=1, \dots, 6$ as a function of the atom number $N$.
    $l_\text{opt}$ increases as $N$ increases.
    We also plot the saturation value in the thermodynamic limit as dashed lines (Eq.~\eqref{eq:l_opt_thermodynamic_limit}).
    $l_\text{opt}$ approaches the saturation value as $N$ increases.
    The fidelities $F(\theta_\text{opt})$ are more than 0.988 for all the data points.
    (b) The measurement probability of $n$ photons $P(n)$ as a function of $N$.
    $P(n)$ increases as $N$ increases for small $N$, while $P(n)$ starts to decrease for large $N$.
    Here, we set $g=g_c$.
  } \label{fig:N_atom_dependence}
\end{figure}

Finally, we investigate the dependence of $\ket{\psi_n}$ on the atom number $N$.
Figure~\hyperref[fig:N_atom_dependence]{4(a)} shows the $N$-dependence of $l_\text{opt}$ for $n=1, \dots, 6$ and $g=g_c$.
$l_\text{opt}$ increases as $N$ increases.
However, $l_\text{opt}$ saturates in the thermodynamic limit $N \to \infty$ as explained in the next section~\ref{sec:thermodynamic_limit}.
We show the saturation value of $l_\text{opt}$ as dashed lines in Fig.~\hyperref[fig:N_atom_dependence]{4(a)}, and we observe that $l_\text{opt}$ approaches the value.

The measurement probability also depends on the atom number $N$.
Figure~\hyperref[fig:N_atom_dependence]{4(b)} shows the $N$ dependence of $P(n)$ for $n=1, \dots, 6$ and $g=g_c$.
We find that $P(n)$ increases as $N$ increases for small $N$.
However, $P(n)$ for $n=1, 2$ starts to decrease for large $N$.
As explained in the next section~\ref{sec:thermodynamic_limit}, $P(n)$ for $n=3, 4, 5, 6$ also starts to decrease at sufficiently large $N$.

\section{Thermodynamic-limit analysis} \label{sec:thermodynamic_limit}

\begin{figure}[t!]
  \includegraphics[width=0.48\textwidth]{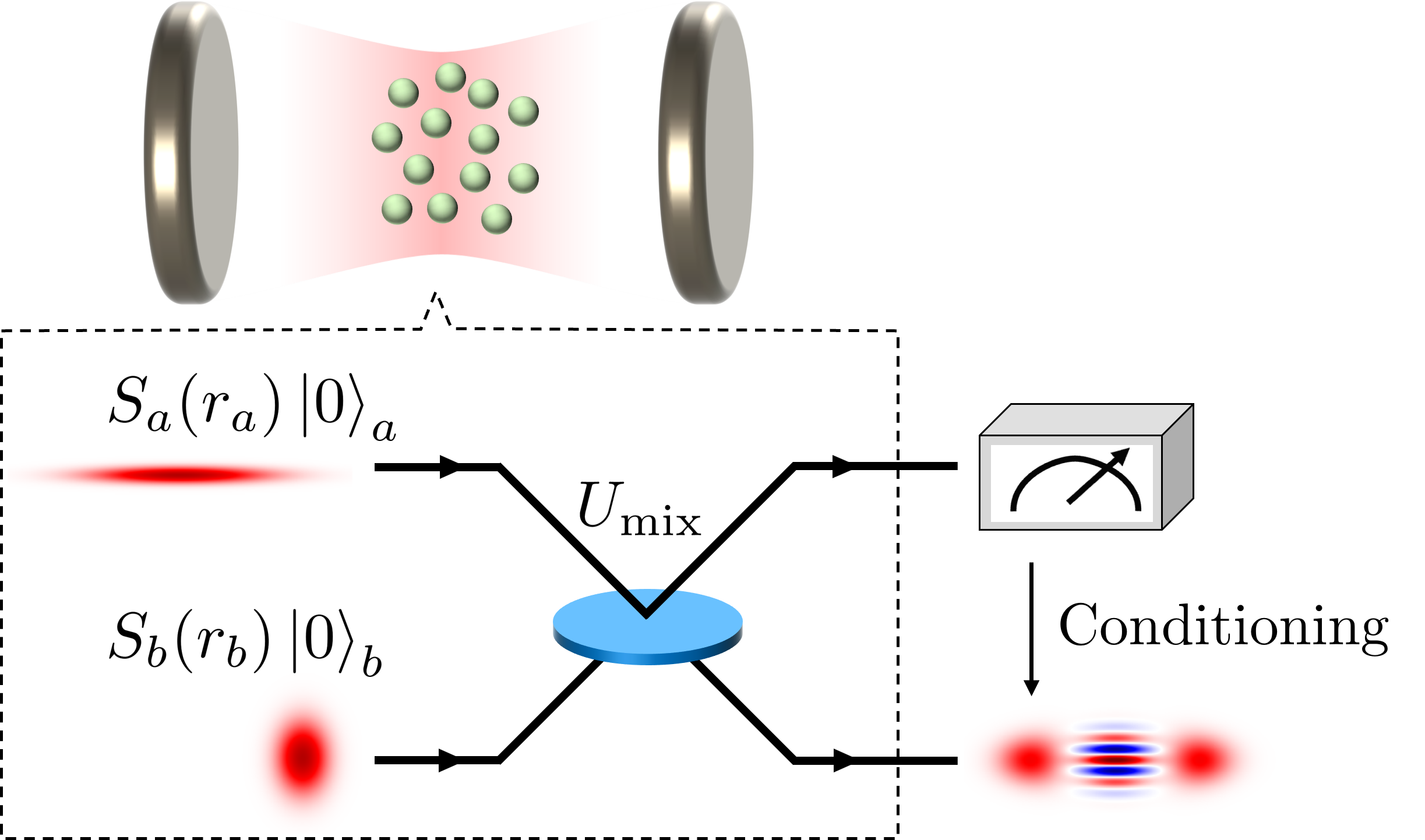}
  \caption{
    The interpretation of the SCS generation protocol from the perspective of the generalized photon subtraction protocol in quantum optics~\cite{takase_Generation_2021}.
    Near the critical point of the superradiant phase transition, the ground state prepares the beam-split squeezed states.
    One mode corresponds to the cavity field, and the other mode corresponds to the collective spin excitation of the atomic ensemble.
    The photon-number measurement on the cavity field can conditionally prepare a SCS similar to the quantum optical case.
  } \label{fig:GPS}
\end{figure}

In this section, we analyze the system in the thermodynamic limit $N \to \infty$ to gain analytical insight into the mechanism for SCS generation.
In this limit, the Dicke model can be solved analytically using the Holstein-Primakoff transformation.
The Holstein-Primakoff transformation maps the collective spin operators to bosonic operators as
\begin{align}
  \hat{J}_+ &= \hat{b}^\dagger \sqrt{N - \hat{b}^\dagger \hat{b}}, \quad \\
  \hat{J}_- &= \sqrt{N - \hat{b}^\dagger \hat{b}} \hat{b}, \quad \\
  \hat{J}_z &= \hat{b}^\dagger \hat{b} - N/2, 
\end{align}
where $\hat{b}^\dagger$ and $\hat{b}$ are bosonic creation and annihilation operators satisfying $[\hat{b}, \hat{b}^\dagger] = 1$.
By substituting these expressions into the Hamiltonian Eq.~\eqref{eq:Dicke_Hamiltonian}, we obtain the following effective Hamiltonian:
\begin{align}
  \hat{H} = \omega \hat{a}^\dagger \hat{a} + \omega \hat{b}^\dagger \hat{b} + g (\hat{a}^\dagger + \hat{a})(\hat{b}^\dagger + \hat{b}) + \order{\frac{1}{N}}. \label{eq:effective_Hamiltonian}
\end{align}
For $g < g_c$, we can neglect the terms of order $1/N$ due to the absence of macroscopic excitation of the atoms.
Therefore, this effective Hamiltonian describes two coupled harmonic oscillators.

The effective Hamiltonian can be diagonalized by performing a unitary transformation.
The unitary operation is a combination of squeezing and mixing operations~\cite{hong-yi_Density_1992,zhou_Quantum_2020,mirkhalaf_Frequency_2025} as
\begin{align}
  \hat{U}^{\dagger}\hat{H}\hat{U} &= \Omega_- \hat{a}^{\dagger}\hat{a} + \Omega_+ \hat{b}^{\dagger}\hat{b} + \text{const.}, \\
  \hat{U} &= \hat{U}_\text{mix} \hat{S}_b(r_b) \hat{S}_a(r_a), \\
  \hat{U}_\text{mix} &= \exp(\frac{\pi}{4}(\hat{a}^\dagger \hat{b} - \hat{a} \hat{b}^\dagger)), \\
  \hat{S}_a(r_a) &= \exp(\frac{r_a}{2}(\hat{a}^2 - \hat{a}^{\dagger 2})), \\
  \hat{S}_b(r_b) &= \exp(\frac{r_b}{2}(\hat{b}^2 - \hat{b}^{\dagger 2})), 
\end{align}
where $\Omega_\pm = \omega\sqrt{1 \pm g/g_c}$, and $r_a = \frac{1}{2}\ln(\Omega_-/\omega)$ and $r_b = \frac{1}{2}\ln(\Omega_+/\omega)$ are the squeezing parameters.
$\hat{S}_a(r_a)$ and $\hat{S}_b(r_b)$ are single-mode squeezing operations for the cavity field and the collective spin excitation, respectively.
$\hat{U}_\text{mix}$ mixes the two modes, and it is equivalent to a beam-splitter operation.
Therefore, the ground state of the system can be written as
\begin{align}
  \ket{G} = \hat{U}_\text{mix} \left(\hat{S}_a(r_a) \ket{0}_a \otimes \hat{S}_b(r_b) \ket{0}_b \right),
\end{align}
where $\ket{0}_a$ and $\ket{0}_b$ are the vacuum states of the bosonic modes $a$ and $b$, respectively.
Approaching the critical point $g \rightarrow g_c$, the squeezing parameters become $r_a \rightarrow -\infty$ and $r_b \rightarrow \frac{1}{4}\ln(2)$.
The diverging behavior of $r_a$ reflects the gap closing behavior of $\Omega_-$ near the critical point.

We point out that the ground state $\ket{G}$ is equivalent to a pair of squeezed states mixed by a beam splitter, which is used in the generalized photon subtraction protocol in quantum optics~\cite{takase_Generation_2021}.
In the generalized photon subtraction, we prepare two squeezed vacuum states of light.
Then, we inject the squeezed vacuum states into a beam splitter, which creates entanglement between the two modes.
Finally, we perform the photon-number measurement on one mode, which conditionally prepares an optical cat state in the other mode.
Therefore, the underlying mechanism for our SCS generation protocol is closely related to the generalized photon subtraction protocol in quantum optics.
Regarding the collective spin excitation as the bosonic mode, the ground state of the Dicke model naturally realizes a pair of squeezed modes mixed by an effective beam splitter as illustrated in Fig.~\hyperref[fig:GPS]{5}.

To obtain an explicit analytical expression of $\ket{\psi_n}$, we rewrite the ground state $\ket{G}$ of the following form:
\begin{align}
  \ket{G} \!=\! \frac{\exp\left(-\frac{\tanh r_+}{2}(\hat{a}^{\dagger}\hat{a}^{\dagger} \!+\! \hat{b}^{\dagger}\hat{b}^{\dagger}) \!+\! \tanh r_- \hat{a}^\dagger \hat{b}^\dagger\right)}{\sqrt{\cosh r_a \cosh r_b}}  \ket{0}_a \ket{0}_b, \label{eq:ground_state_expansion}
\end{align}
where $\tanh r_\pm = \frac{1}{2}(\tanh r_a \pm \tanh r_b)$.
By expanding the exponential factor and projecting onto the $n$-photons component, we get concrete expressions for $\ket{\psi_n}$.
For example, the $n=1$ and $n=2$ cases are given by
\begin{align}
  \ket{\psi_1} &\propto \hat{b} \hat{S}_b(r_+) \ket{0}_b, \\
  \ket{\psi_2} &\propto \left(\frac{\tanh^2 r_-}{\tanh r_+}\hat{b}^2 - \tanh r_a \tanh r_b\right)\hat{S}_b(r_+) \ket{0}_b.
\end{align}
$\ket{\psi_1}$ is a single-boson subtracted squeezed state, while $\ket{\psi_2}$ includes a two-boson subtracted squeezed state.
In general, $\ket{\psi_n}$ includes the $n$-boson subtracted squeezed state  (see Appendix~\ref{sec:detail_ground_state} for the details of the derivation).
The $n$-boson subtraction creates an interference pattern in the Wigner function.

\begin{figure}[t!]
  \includegraphics[width=0.48\textwidth]{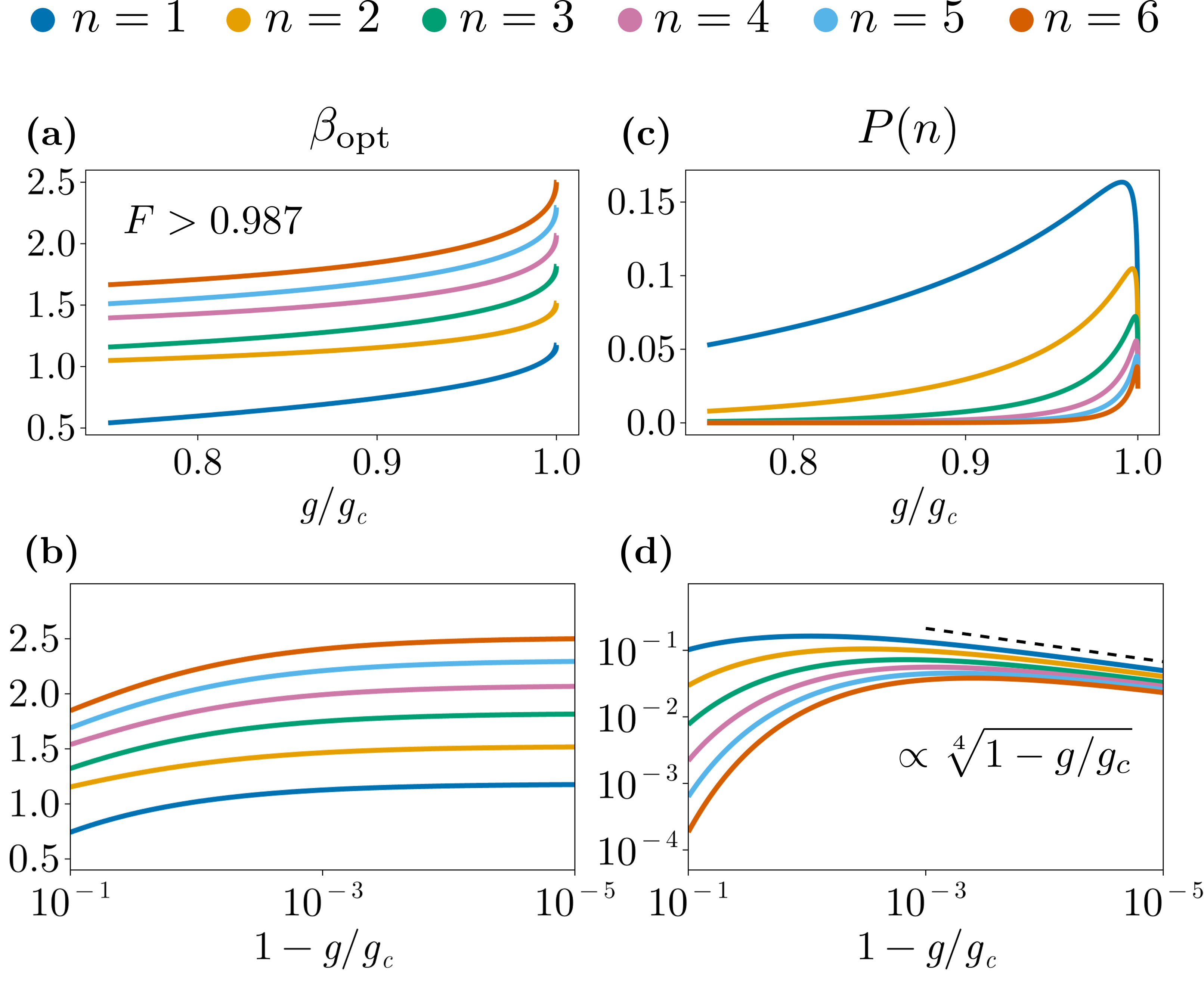}
  \caption{
    (a) (b) The optimal amplitude $\beta_\text{opt}$ as a function of (a) the coupling strength $g$ and (b) $1-g/g_c$ for $n=1, \dots, 6$.
    $\beta_\text{opt}$ grows significantly as $g$ increases, but it saturates when $g$ becomes close to $g_c$.
    The fidelities $F(\beta_\text{opt})$ are more than 0.987 for all the data points.
    (c) (d) The measurement probability of $n$ photons $P(n)$ as a function of (c) $g$ and (d) $1-g/g_c$ for $n=1, \dots, 6$.
    $P(n)$ increases as $g$ increases, but it starts to decrease when $g$ approaches $g_c$.
    The decreasing behavior of $P(n)$ is proportional to $\sqrt[4]{1-g/g_c}$.
  } \label{fig:g_critical}
\end{figure}

In order to quantify the size of the SCS in the thermodynamic limit, we consider the bosonic cat state defined as $\ket{\text{cat}(\beta)}_b \propto \ket{\beta}_b \pm \ket{-\beta}_b$.
$\ket{\beta}_b$ is the bosonic coherent state.
We calculate the optimal amplitude $\beta_\text{opt}$ by maximizing the fidelity $F(\beta) = |_b\bra{\text{cat}(\beta)}\ket{\psi_n}|^2$ as
\begin{align}
  \beta_\text{opt} = \text{argmax}_{\beta} F(\beta).
\end{align}
We note that $2\beta_\text{opt}$ gives the SCS size in the thermodynamic limit as follows.
The quantum fluctuation of $(\hat{b}^{\dagger} + \hat{b})/2$ in the bosonic coherent state is $1/2$, which corresponds to the quantum fluctuation $\sqrt{J/2}$ in the coherent spin state.
By dividing $\beta_\text{opt}$ by the quantum fluctuation $1/2$, we get $l_\text{opt}$ in the thermodynamic limit as
\begin{align}
  \lim_{N \rightarrow \infty} l_\text{opt} = 2\beta_\text{opt}. \label{eq:l_opt_thermodynamic_limit}
\end{align}
The dashed lines in Fig.~\hyperref[fig:N_atom_dependence]{4(a)} represent $2\beta_\text{opt}$ for $g \rightarrow g_c$.
It is observed that $l_\text{opt}$ for finite $N$ approaches $2\beta_\text{opt}$ as $N$ increases.

In order to see the cat state behavior near the critical point in the thermodynamic limit, we calculate the $g$-dependence of $\beta_\text{opt}$.
Figure~\hyperref[fig:g_critical]{6(a)} shows the $g$-dependence of the optimal amplitude $\beta_\text{opt}$ for $n=1, \dots, 6$.
Near the critical point $g=g_c$, the size of the cat state $\beta_\text{opt}$ grows significantly.
This behavior is consistent with the finite-size results as shown in Fig.~\hyperref[fig:g_dependence]{3(a)}.
Moreover, we find that $\beta_\text{opt}$ saturates when $g$ becomes close to $g_c$.
We take $1-g/g_c$ as the horizontal axis in Fig.~\hyperref[fig:g_critical]{6(b)} to see the saturation.
It is apparent that $\beta_\text{opt}$ approaches a constant value when $g \rightarrow g_c$.

We also investigate the measurement probability to detect $n$ photons near the critical point in the thermodynamic limit.
Figure~\hyperref[fig:g_critical]{6(c)} shows the $g$-dependence of $P(n)$ for $n=1, \dots, 6$.
$P(n)$ increases exponentially as $g$ increases, which is consistent with the finite-size results in Fig.~\hyperref[fig:g_dependence]{3(b)}.
However, we find that $P(n)$ starts to decrease near the critical point, which is not observed in the $N=200$ data in Fig.~\hyperref[fig:g_dependence]{3(b)}.
This can be explained as follows.
The normalization prefactor in Eq.~\eqref{eq:ground_state_expansion} contains $1/\sqrt{\cosh r_a} \propto \sqrt[8]{1-g/g_c}$ as $g \rightarrow g_c$.
Consequently, the fixed-$n$ probability scales as $P(n) \propto 1/\cosh r_a \propto \sqrt[4]{1-g/g_c}$ near the critical point as shown in Fig.~\hyperref[fig:g_critical]{6(d)}.
The measurement probability for fixed $n$ approaches zero, which is consistent with the observations in Fig.~\hyperref[fig:N_atom_dependence]{4(b)}.
We note that the measurement probability of the larger photon number is relatively enhanced because the measurement probability of large photon number $n$ becomes closer to that of small photon number $n$.

\section{Discussion and Conclusion} \label{sec:discussion_conclusion}
We now discuss the relation to previous works on the SCS generation and possible experimental implementation of the proposed measurement-based protocol.
We first compare our measurement-based scheme with the deterministic scheme for generating SCSs.
The one-axis twisting~\cite{kitagawa_Squeezed_1993} generates a cat state at a certain time~\cite{groiseau_Generation_2021,hotter_Conditional_2025}.
However, this scheme requires strong all-to-all interactions between each spin and preparation before dephasing and dissipation.
The measurement-based scheme provides an alternative strategy for generating SCSs without all-to-all atom interactions and fine-tuning of the evolution time.
Our scheme generates SCSs with a single-shot number-resolved photon measurement to the ground state whenever the measurement outcome is non-zero.

We also comment on the experimental feasibility of the present SCS generation protocol.
The Dicke model and superradiant phase transition have been experimentally realized in Bose-Einstein condensates in cavities~\cite{baumann_Dicke_2010,mivehvar_Cavity_2021}.
In these experiments, the superradiant phase transition has been observed by measuring the cavity field leaking out of the cavity.
On the other hand, our protocol requires the photon-number measurement of the intracavity field.
The intracavity photon-number measurement has also been achieved in cavity systems~\cite{guerlin_Progressive_2007,gleyzes_Quantum_2007}.
Therefore, combining these experimental platforms can be a promising route to implementing our SCS generation protocol.

An important caveat for experimental implementation is that superradiant phase transitions in cavity-QED experiments are often realized as driven-dissipative nonequilibrium transitions~\cite{mivehvar_Cavity_2021}, whereas our analysis assumes the ground state of the equilibrium Dicke model. 
It is therefore important to extend the present protocol to the steady state of a driven-dissipative Dicke model~\cite{kirton_Introduction_2019} and to clarify the robustness of the heralded SCSs against cavity loss and atomic decay.

In conclusion, we proposed a method for generating SCSs in atomic ensembles by utilizing the criticality of the superradiant phase transition.
By measuring the photon number of the cavity field near the critical point, we can conditionally prepare high-fidelity SCSs in the atomic ensemble.
We find that the size of the SCSs is significantly enhanced when the system approaches the critical point.
The criticality also enhances the measurement probability of larger photon number, thereby leading to larger SCSs.
We show that the mechanism for our SCS generation protocol can be regarded as an atom-light version of the generalized photon subtraction protocol.
Our work opens a route to generating non-classical states in atomic ensembles exploiting the criticality of quantum phase transitions.

\acknowledgments
We are grateful to N. Tsuji and M. Ueda for helpful discussions.
T.N. is supported by WINGS-MERIT of the University of Tokyo and JST SPRING (Grant No.~JPMJSP2108).
S.I. is supported by JST FOREST (Grant No. JPMJFR2131) and JSPS KAKENHI (Grants Nos. JP25K17343 and JP26K21749).
K.T. is supported by JST PRESTO (Grants Nos. JPMJPR2256 and JPMJPR2596) and JSPS KAKENHI (Grants Nos. JP23K17664, JP25K17312,  JP26H00385, and JP26H00382).

\appendix

\section{Off-resonant case} \label{off_resonant}

\begin{figure}[t!]
  \includegraphics[width=0.48\textwidth]{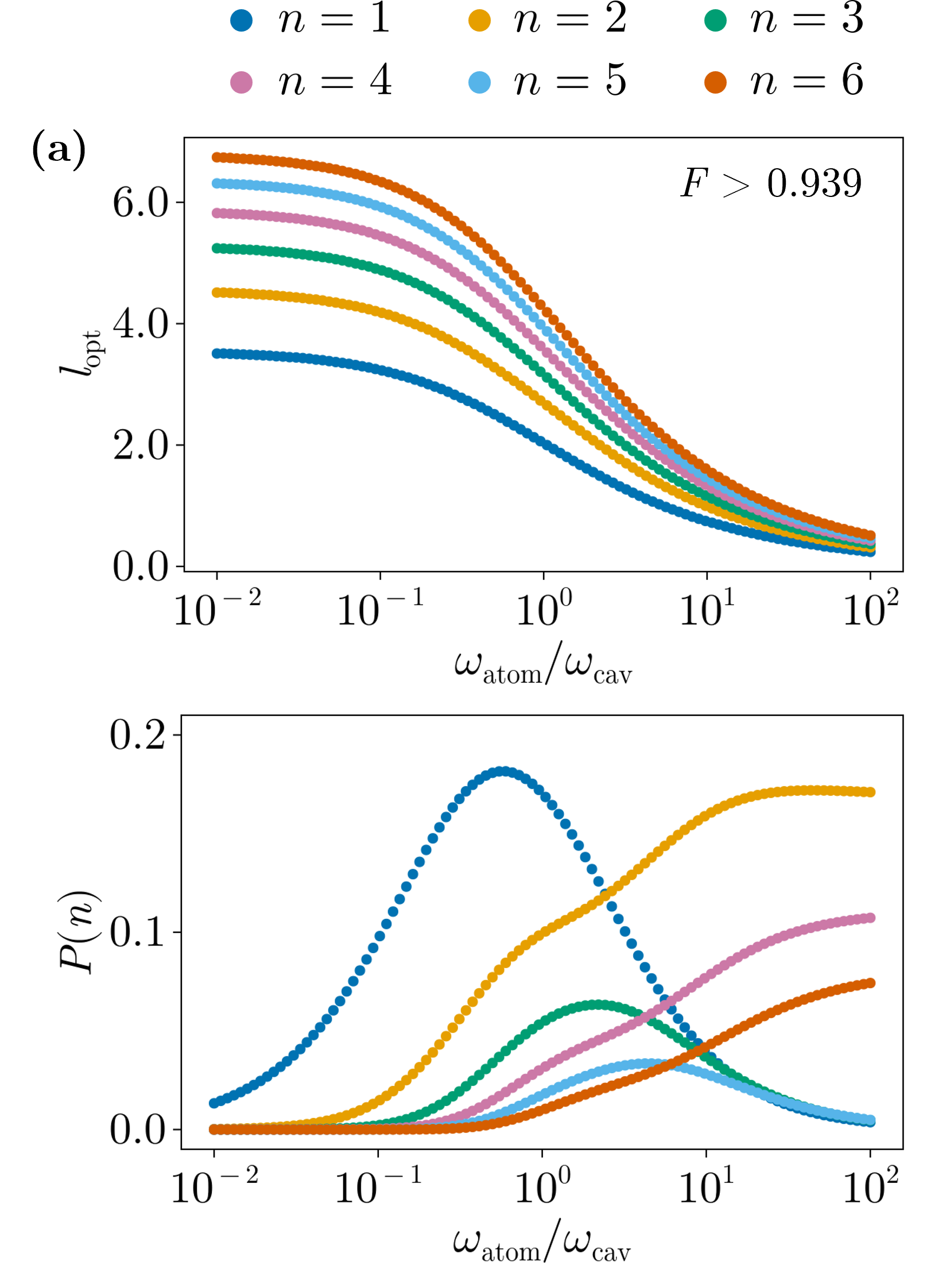}
  \caption{
    (a) The size of the SCS $l_\text{opt}$ for $n=1, \dots, 6$ as a function of the atom-cavity frequency ratio $\omega_\text{atom}/\omega_\text{cav}$.
    $l_\text{opt}$ is significantly enhanced when $\omega_\text{atom} \ll \omega_\text{cav}$.
    The fidelities $F(\theta_\text{opt})$ are more than 0.939 for all the data points.
    (b) The measurement probability of $n$ photons $P(n)$ as a function of $\omega_\text{atom}/\omega_\text{cav}$.
    $P(n)$ is enhanced when $\omega_\text{atom} \gg \omega_\text{cav}$ for even $n$, while $P(n)$ has a maximum around $\omega_\text{atom} \sim \omega_\text{cav}$ for odd $n$.
    Here, we set $N=200$ and $g=g_c$.
  } \label{fig:frequency_dependence}
\end{figure}

In this appendix, we present calculations for the off-resonant case $\omega_\text{cav} \neq \omega_\text{atom}$ at the critical point $g=g_c$.
Figure~\hyperref[fig:frequency_dependence]{7(a)} shows the dependence of the SCS size $l_\text{opt}$ on the atom-cavity frequency ratio $\omega_\text{atom}/\omega_\text{cav}$ for $n=1, \dots, 6$.
The data show that $l_\text{opt}$ significantly increases when $\omega_\text{atom} \ll \omega_\text{cav}$. 
Therefore, decreasing the atom frequency relative to the cavity frequency can be a strategy to enhance the size of the SCS.

However, the $n$-photon measurement probability $P(n)$ is small when $\omega_\text{atom} \ll \omega_\text{cav}$ as shown in Fig.~\hyperref[fig:frequency_dependence]{7(b)}.
$P(n)$ is enhanced when $\omega_\text{atom} \gg \omega_\text{cav}$ for even $n$, while $P(n)$ has a maximum around $\omega_\text{atom} \sim \omega_\text{cav}$ for odd $n$.
Therefore, $\omega_\text{atom} \sim \omega_\text{cav}$ is a good choice for balancing the SCS size and the measurement probability.

We note that the fidelities between $\ket{\psi_n}$ and $\ket{\text{cat}(\theta_\text{opt})}$ remain above 0.939, although this lower bound is smaller than that in the resonant case $\omega_\text{atom} = \omega_\text{cav}$ in the main text.
The fidelities exceed 0.995 when $\omega_\text{atom} \geq \omega_\text{cav}$, while the fidelities monotonically decrease as $\omega_\text{atom}/\omega_\text{cav}$ decreases when $\omega_\text{atom} < \omega_\text{cav}$.
Thus, the SCSs in the resonant case are closer to the ideal SCSs than those in the case of $\omega_\text{atom} < \omega_\text{cav}$.

\section{Numerical convergence} \label{sec:numerical_convergence}

\begin{figure}[t!]
  \includegraphics[width=0.48\textwidth]{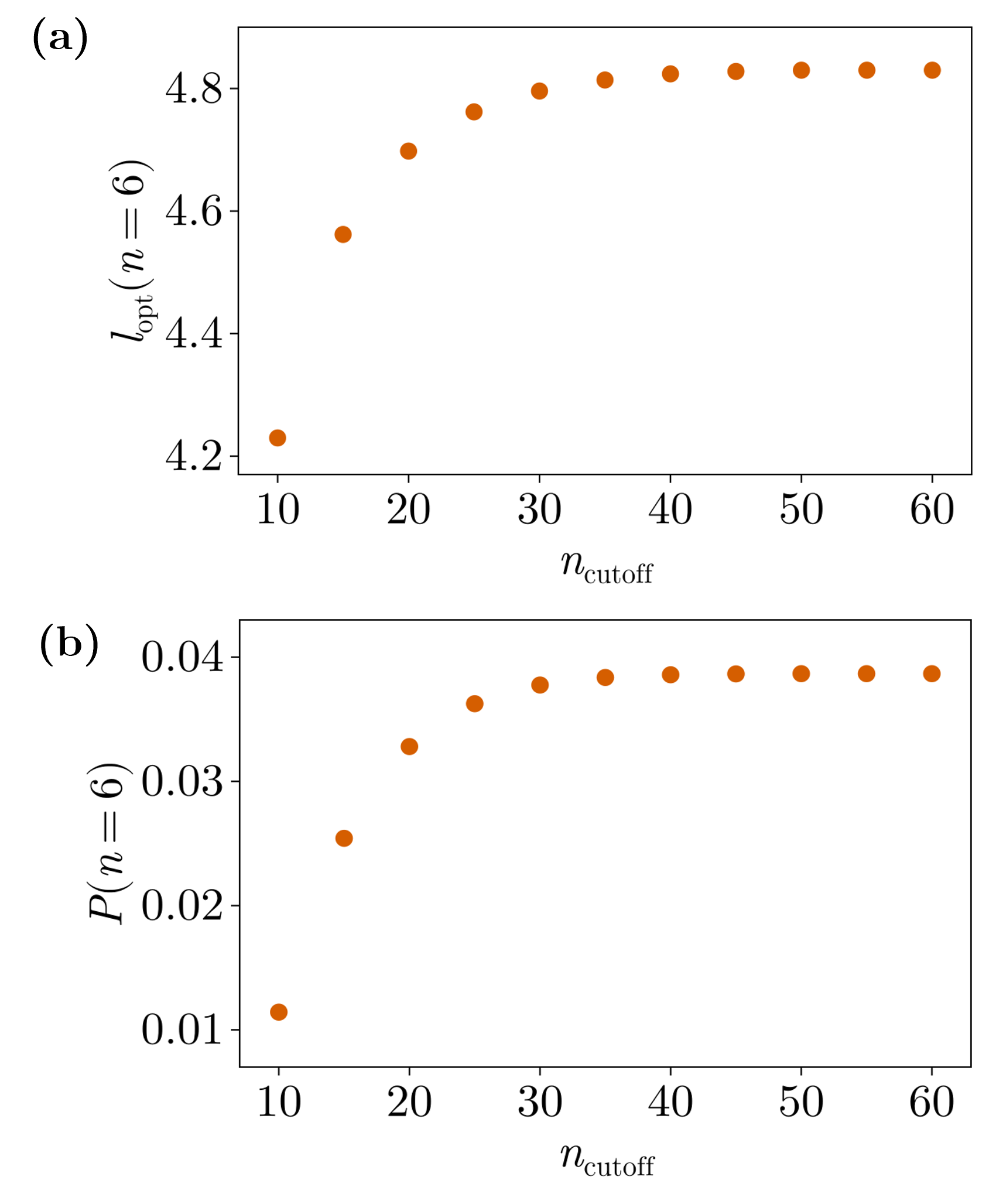}
  \caption{
    (a) The size of the SCS $l_\text{opt}$ and (b) the measurement probability of $n$ photons $P(n)$ as a function of the photon-number cutoff $n_\text{cutoff}$.
    We set $n=6$, $N=10^{4}$, and $g=g_c$.
    $l_\text{opt}$ and $P(n)$ converge when $n_\text{cutoff} \geq 50$.
  } \label{fig:cutoff_dependence}
\end{figure}

In this appendix, we confirm the numerical convergence of our results.
We perform the exact diagonalization of the Dicke Hamiltonian by truncating the maximum photon number $n_\text{cutoff}$.
Figure~\hyperref[fig:cutoff_dependence]{8(a)} and \hyperref[fig:cutoff_dependence]{8(b)} show the $n_\text{cutoff}$-dependence of $l_\text{opt}$ and $P(n)$ for $n=6$, $N=10^{4}$, and $g=g_c$.
$n=6$ and $N=10^{4}$ are the largest photon number and atom number in our numerical calculations, respectively.
We confirm that $l_\text{opt}$ and $P(n)$ converge when $n_\text{cutoff} \geq 50$.

\section{Details of the ground state expansion} \label{sec:detail_ground_state}

In this appendix, we derive the ground state expansion in Eq.~\eqref{eq:ground_state_expansion} and the expressions of the $n$-photon heralded state $\ket{\psi_n}$.
We expand $\ket{G} = \hat{U}_\text{mix} \hat{S}_a(r_a) \hat{S}_b(r_b) \ket{0}_a \ket{0}_b$ using the disentangling formula for the squeezing operator given by
\begin{align}
  \hat{S}(r)\ket{0} = \frac{1}{\sqrt{\cosh r}} \exp\left(-\frac{\tanh r}{2} \hat{a}^{\dagger 2}\right) \ket{0}. \label{eq:disentangling_formula}
\end{align}
The denominator $\sqrt{\cosh r_a \cosh r_b}$ in Eq.~\eqref{eq:ground_state_expansion} comes from this relation.
$\hat{U}_\text{mix}$ acts on the exponential factor as
\begin{align}
  \ket{G} \propto& \hat{U}_\text{mix} \exp(-\frac{\tanh r_a}{2} \hat{a}^{\dagger 2} - \frac{\tanh r_b}{2} \hat{b}^{\dagger 2}) \ket{0}_a \ket{0}_b, \nonumber \\
  =& \exp(-\frac{\tanh r_a}{2} \hat{a'}^{\dagger 2} - \frac{\tanh r_b}{2} \hat{b'}^{\dagger 2} ) \ket{0}_a \ket{0}_b, 
\end{align}
where 
\begin{align}
  \hat{a'}^{\dagger} =& \hat{U}_\text{mix} \hat{a}^{\dagger} \hat{U}_\text{mix}^\dagger \nonumber \\
  =& \frac{1}{\sqrt{2}}(\hat{a}^\dagger - \hat{b}^\dagger), \\
  \hat{b'}^{\dagger} =& \hat{U}_\text{mix} \hat{b}^{\dagger} \hat{U}_\text{mix}^\dagger \nonumber \\
  =& \frac{1}{\sqrt{2}}(\hat{a}^\dagger + \hat{b}^\dagger).
\end{align}
Here, we use the relation $\hat{U}_\text{mix} \ket{0}_a \ket{0}_b = \ket{0}_a \ket{0}_b$.
By simplifying the exponent, we obtain the expression in Eq.~\eqref{eq:ground_state_expansion}.

Next, we derive the concrete expressions of $\ket{\psi_n}$.
We expand the exponential factor in Eq.~\eqref{eq:ground_state_expansion} as
\begin{align}
  \ket{G} \!\propto\! \!\sum_{k,l,m}^{\infty}\! \frac{(-\frac{\tanh r_+}{2})^{k+l}(\tanh r_-)^m \hat{a}^{\dagger 2k+m}\hat{b}^{\dagger 2l+m} \ket{0}_a \ket{0}_b  }{k! l! m!}
\end{align}
By projecting onto the $\ket{n}_a$ component, we choose the pairs of integers $(k,m)$ that satisfies $2k+m=n$.
For example, the $n=1$ case corresponds to $(k,m)=(0,1)$, while the $n=2$ case corresponds to $(k,m)=(1,0)$ and $(0,2)$. 
For a fixed pair $(k,m)$, $\ket{\psi_n}$ is proportional to the following state:
\begin{align}
  \ket{\psi_n} \propto& \hat{b}^{\dagger m} \sum_{l=0}^{\infty} \frac{(-\frac{\tanh r_+}{2} \hat{b}^{\dagger 2})^l}{l!} \ket{0}_b \nonumber \\
  \propto& \hat{b}^{\dagger m} \hat{S}_b(r_+) \ket{0}_b. \label{eq:psi_n_expansion}
\end{align}
Here, we again use the disentangling formula~\eqref{eq:disentangling_formula}.
Using the relation 
\begin{align}
  \hat{b}\hat{S}_b(r)\ket{0}_b = (-\tanh r) \hat{b}^\dagger \hat{S}_b(r)\ket{0}_b,
\end{align}
Eq.~\eqref{eq:psi_n_expansion} is rewritten as a $m$-boson subtracted squeezed state.
Therefore, $\ket{\psi_n}$ is a superposition of the $m$-boson subtracted squeezed states with $m = n, n-2, n-4, \dots$.

\bibliography{reference}

@article{schrodinger_Gegenwartige_1935,
  title = {{Die gegenw\"artige Situation in der Quantenmechanik}},
  author = {Schr{\"o}dinger, E.},
  year = 1935,
  month = dec,
  journal = {Naturwissenschaften},
  volume = {23},
  number = {49},
  pages = {823--828},
  doi = {10.1007/BF01491914}
}

@article{deleglise_Reconstruction_2008,
  title = {Reconstruction of Non-Classical Cavity Field States with Snapshots of Their Decoherence},
  author = {Del{\'e}glise, Samuel and Dotsenko, Igor and Sayrin, Cl{\'e}ment and Bernu, Julien and Brune, Michel and Raimond, Jean-Michel and Haroche, Serge},
  year = 2008,
  month = sep,
  journal = {Nature},
  volume = {455},
  number = {7212},
  pages = {510--514},
  doi = {10.1038/nature07288}
}

@article{munro_Weakforce_2002,
  title = {Weak-Force Detection with Superposed Coherent States},
  author = {Munro, W. J. and Nemoto, K. and Milburn, G. J. and Braunstein, S. L.},
  year = 2002,
  month = aug,
  volume = {66},
  number = {2},
  pages = {023819},
  doi = {10.1103/PhysRevA.66.023819},
  journal = {Phys. Rev. A}
}

@article{joo_Quantum_2011,
  title = {Quantum {{Metrology}} with {{Entangled Coherent States}}},
  author = {Joo, Jaewoo and Munro, William J. and Spiller, Timothy P.},
  year = 2011,
  month = aug,
  volume = {107},
  number = {8},
  pages = {083601},
  doi = {10.1103/PhysRevLett.107.083601},
  journal = {Phys. Rev. Lett.}
}

@article{huang_Quantum_2015,
  title = {Quantum Metrology with Spin Cat States under Dissipation},
  author = {Huang, Jiahao and Qin, Xizhou and Zhong, Honghua and Ke, Yongguan and Lee, Chaohong},
  year = 2015,
  month = dec,
  volume = {5},
  number = {1},
  pages = {17894},
  doi = {10.1038/srep17894},
  journal = {Sci Rep}
}

@article{ralph_Quantum_2003,
  title = {Quantum Computation with Optical Coherent States},
  author = {Ralph, T. C. and Gilchrist, A. and Milburn, G. J. and Munro, W. J. and Glancy, S.},
  year = 2003,
  month = oct,
  volume = {68},
  number = {4},
  pages = {042319},
  doi = {10.1103/PhysRevA.68.042319},
  journal = {Phys. Rev. A}
}

@article{pezze_Quantum_2018,
  title = {Quantum Metrology with Nonclassical States of Atomic Ensembles},
  author = {Pezz{\`e}, Luca and Smerzi, Augusto and Oberthaler, Markus K. and Schmied, Roman and Treutlein, Philipp},
  year = 2018,
  month = sep,
  volume = {90},
  number = {3},
  pages = {035005},
  doi = {10.1103/RevModPhys.90.035005},
  journal = {Rev. Mod. Phys.}
}

@article{cochrane_Macroscopically_1999,
  title = {Macroscopically Distinct Quantum-Superposition States as a Bosonic Code for Amplitude Damping},
  author = {Cochrane, P. T. and Milburn, G. J. and Munro, W. J.},
  year = 1999,
  month = apr,
  volume = {59},
  number = {4},
  pages = {2631--2634},
  doi = {10.1103/PhysRevA.59.2631},
  journal = {Phys. Rev. A}
}

@article{ofek_Extending_2016,
  title = {Extending the Lifetime of a Quantum Bit with Error Correction in Superconducting Circuits},
  author = {Ofek, Nissim and Petrenko, Andrei and Heeres, Reinier and Reinhold, Philip and Leghtas, Zaki and Vlastakis, Brian and Liu, Yehan and Frunzio, Luigi and Girvin, S. M. and Jiang, L. and Mirrahimi, Mazyar and Devoret, M. H. and Schoelkopf, R. J.},
  year = 2016,
  month = aug,
  journal = {Nature},
  volume = {536},
  number = {7617},
  pages = {441--445},
  doi = {10.1038/nature18949}
}

@article{lescanne_Exponential_2020,
  title = {Exponential Suppression of Bit-Flips in a Qubit Encoded in an Oscillator},
  author = {Lescanne, Rapha{\"e}l and Villiers, Marius and Peronnin, Th{\'e}au and Sarlette, Alain and Delbecq, Matthieu and Huard, Benjamin and Kontos, Takis and Mirrahimi, Mazyar and Leghtas, Zaki},
  year = 2020,
  month = may,
  volume = {16},
  number = {5},
  pages = {509--513},
  doi = {10.1038/s41567-020-0824-x},
  journal = {Nat. Phys.}
}

@article{frowis_Macroscopic_2018,
  title = {Macroscopic Quantum States: {{Measures}}, Fragility, and Implementations},
  author = {Fr{\"o}wis, Florian and Sekatski, Pavel and D{\"u}r, Wolfgang and Gisin, Nicolas and Sangouard, Nicolas},
  year = 2018,
  month = may,
  volume = {90},
  number = {2},
  pages = {025004},
  doi = {10.1103/RevModPhys.90.025004},
  journal = {Rev. Mod. Phys.}
}

@article{dakna_Generating_1997,
   title = {Generating {{Schr\"odinger}}-Cat-like States by Means of Conditional Measurements on a Beam Splitter},
   author = {Dakna, M. and Anhut, T. and Opatrn{\'y}, T. and Kn{\"o}ll, L. and Welsch, D.-G.},
   year = 1997,
   month = apr,
   volume = {55},
   number = {4},
   pages = {3184--3194},
   doi = {10.1103/PhysRevA.55.3184},
   journal = {Phys. Rev. A}
}

@article{ourjoumtsev_Generating_2006,
  title = {Generating {{Optical Schr\"odinger Kittens}} for {{Quantum Information Processing}}},
  author = {Ourjoumtsev, Alexei and {Tualle-Brouri}, Rosa and Laurat, Julien and Grangier, Philippe},
  year = 2006,
  month = apr,
  journal = {Science},
  volume = {312},
  number = {5770},
  pages = {83--86},
  doi = {10.1126/science.1122858}
}

@article{neergaard_Generation_2006,
  title = {Generation of a {{Superposition}} of {{Odd Photon Number States}} for {{Quantum Information Networks}}},
  author = {{Neergaard-Nielsen}, J. S. and Nielsen, B. Melholt and Hettich, C. and M{\o}lmer, K. and Polzik, E. S.},
  year = 2006,
  month = aug,
  volume = {97},
  number = {8},
  pages = {083604},
  doi = {10.1103/PhysRevLett.97.083604},
  journal = {Phys. Rev. Lett.}
}

@article{takahashi_Generation_2008,
  title = {Generation of {{Large-Amplitude Coherent-State Superposition}} via {{Ancilla-Assisted Photon Subtraction}}},
  author = {Takahashi, Hiroki and Wakui, Kentaro and Suzuki, Shigenari and Takeoka, Masahiro and Hayasaka, Kazuhiro and Furusawa, Akira and Sasaki, Masahide},
  year = 2008,
  month = dec,
  volume = {101},
  number = {23},
  pages = {233605},
  doi = {10.1103/PhysRevLett.101.233605},
  journal = {Phys. Rev. Lett.}
}

@article{gerrits_Generation_2010,
  title = {Generation of Optical Coherent-State Superpositions by Number-Resolved Photon Subtraction from the Squeezed Vacuum},
  author = {Gerrits, Thomas and Glancy, Scott and Clement, Tracy S. and Calkins, Brice and Lita, Adriana E. and Miller, Aaron J. and Migdall, Alan L. and Nam, Sae Woo and Mirin, Richard P. and Knill, Emanuel},
  year = 2010,
  month = sep,
  volume = {82},
  number = {3},
  pages = {031802},
  doi = {10.1103/PhysRevA.82.031802},
  journal = {Phys. Rev. A}
}

@article{huang_Optical_2015,
  title = {Optical {{Synthesis}} of {{Large-Amplitude Squeezed Coherent-State Superpositions}} with {{Minimal Resources}}},
  author = {Huang, K. and Le Jeannic, H. and Ruaudel, J. and Verma, V. B. and Shaw, M. D. and Marsili, F. and Nam, S. W. and Wu, E and Zeng, H. and Jeong, Y.-C. and Filip, R. and Morin, O. and Laurat, J.},
  year = 2015,
  month = jul,
  volume = {115},
  number = {2},
  pages = {023602},
  doi = {10.1103/PhysRevLett.115.023602},
  journal = {Phys. Rev. Lett.}
}

@article{sychev_Enlargement_2017,
  title = {Enlargement of Optical {{Schr\"odinger}}'s Cat States},
  author = {Sychev, Demid V. and Ulanov, Alexander E. and Pushkina, Anastasia A. and Richards, Matthew W. and Fedorov, Ilya A. and Lvovsky, Alexander I.},
  year = 2017,
  month = jun,
  volume = {11},
  number = {6},
  pages = {379--382},
  doi = {10.1038/nphoton.2017.57},
  journal = {Nature Photon}
}

@article{takase_Generation_2021,
  title = {Generation of Optical {{Schr\"odinger}} Cat States by Generalized Photon Subtraction},
  author = {Takase, Kan and Yoshikawa, Jun-ichi and Asavanant, Warit and Endo, Mamoru and Furusawa, Akira},
  year = 2021,
  month = jan,
  volume = {103},
  number = {1},
  pages = {013710},
  doi = {10.1103/PhysRevA.103.013710},
  journal = {Phys. Rev. A}
}

@article{imai_Macroscopic_2026,
  title = {Macroscopic {{Schr\"odinger}}-Cat States of Nonequilibrium Electrons Induced by Cat-State Optical Driving and Projective Measurements on the Light Field},
  author = {Imai, Shohei},
  year = 2026,
  month = feb,
  volume = {113},
  number = {2},
  pages = {023702},
  doi = {10.1103/kzr9-fxx3},
  journal = {Phys. Rev. A}
}

@article{mcconnell_Entanglement_2015,
  title = {Entanglement with Negative {{Wigner}} Function of Almost 3,000 Atoms Heralded by One Photon},
  author = {McConnell, Robert and Zhang, Hao and Hu, Jiazhong and {\'C}uk, Senka and Vuleti{\'c}, Vladan},
  year = 2015,
  month = mar,
  journal = {Nature},
  volume = {519},
  number = {7544},
  pages = {439--442},
  doi = {10.1038/nature14293}
}

@article{massar_Generating_2003,
  title = {Generating a {{Superposition}} of {{Spin States}} in an {{Atomic Ensemble}}},
  author = {Massar, S. and Polzik, E. S.},
  year = 2003,
  month = aug,
  volume = {91},
  number = {6},
  pages = {060401},
  doi = {10.1103/PhysRevLett.91.060401},
  journal = {Phys. Rev. Lett.}
}

@article{genes_Generating_2006,
  title = {Generating Conditional Atomic Entanglement by Measuring Photon Number in a Single Output Channel},
  author = {Genes, C. and Berman, P. R.},
  year = 2006,
  month = jan,
  volume = {73},
  number = {1},
  pages = {013801},
  doi = {10.1103/PhysRevA.73.013801},
  journal = {Phys. Rev. A}
}

@article{mcconnell_Generating_2013,
  title = {Generating Entangled Spin States for Quantum Metrology by Single-Photon Detection},
  author = {McConnell, Robert and Zhang, Hao and {\'C}uk, Senka and Hu, Jiazhong and {Schleier-Smith}, Monika H. and Vuleti{\'c}, Vladan},
  year = 2013,
  month = dec,
  volume = {88},
  number = {6},
  pages = {063802},
  doi = {10.1103/PhysRevA.88.063802},
  journal = {Phys. Rev. A}
}

@article{pettersson_Lightmediated_2017,
  title = {Light-Mediated Non-{{Gaussian}} Entanglement of Atomic Ensembles},
  author = {Pettersson, Olov and Byrnes, Tim},
  year = 2017,
  month = apr,
  volume = {95},
  number = {4},
  pages = {043817},
  doi = {10.1103/PhysRevA.95.043817},
  journal = {Phys. Rev. A}
}

@article{davis_Painting_2018,
  title = {Painting {{Nonclassical States}} of {{Spin}} or {{Motion}} with {{Shaped Single Photons}}},
  author = {Davis, Emily J. and Wang, Zhaoyou and {Safavi-Naeini}, Amir H. and {Schleier-Smith}, Monika H.},
  year = 2018,
  month = sep,
  volume = {121},
  number = {12},
  pages = {123602},
  doi = {10.1103/PhysRevLett.121.123602},
  journal = {Phys. Rev. Lett.}
}

@article{kitagawa_Squeezed_1993,
  title = {Squeezed Spin States},
  author = {Kitagawa, Masahiro and Ueda, Masahito},
  year = 1993,
  month = jun,
  volume = {47},
  number = {6},
  pages = {5138--5143},
  doi = {10.1103/PhysRevA.47.5138},
  journal = {Phys. Rev. A}
}

@article{groiseau_Generation_2021,
  title = {Generation of Spin Cat States in an Engineered {{Dicke}} Model},
  author = {Groiseau, Caspar and Masson, Stuart J. and Parkins, Scott},
  year = 2021,
  month = nov,
  volume = {104},
  number = {5},
  pages = {053721},
  doi = {10.1103/PhysRevA.104.053721},
  journal = {Phys. Rev. A}
}

@article{hotter_Conditional_2025,
  title = {Conditional {{Entanglement Amplification}} via {{Non-Hermitian Superradiant Dynamics}}},
  author = {Hotter, Christoph and Kosior, Arkadiusz and Ritsch, Helmut and Gietka, Karol},
  year = 2025,
  month = jun,
  volume = {134},
  number = {23},
  pages = {233601},
  doi = {10.1103/w377-f9mx},
  journal = {Phys. Rev. Lett.}
}

@article{hepp_Equilibrium_1973,
  title = {Equilibrium {{Statistical Mechanics}} of {{Matter Interacting}} with the {{Quantized Radiation Field}}},
  author = {Hepp, Klaus and Lieb, Elliott H.},
  year = 1973,
  month = nov,
  volume = {8},
  number = {5},
  pages = {2517--2525},
  doi = {10.1103/PhysRevA.8.2517},
  journal = {Phys. Rev. A}
}

@article{wang_Phase_1973,
  title = {Phase {{Transition}} in the {{Dicke Model}} of {{Superradiance}}},
  author = {Wang, Y. K. and Hioe, F. T.},
  year = 1973,
  month = mar,
  volume = {7},
  number = {3},
  pages = {831--836},
  doi = {10.1103/PhysRevA.7.831},
  journal = {Phys. Rev. A}
}

@article{lambert_Entanglement_2004,
  title = {Entanglement and the {{Phase Transition}} in {{Single-Mode Superradiance}}},
  author = {Lambert, Neill and Emary, Clive and Brandes, Tobias},
  year = 2004,
  month = feb,
  volume = {92},
  number = {7},
  pages = {073602},
  doi = {10.1103/PhysRevLett.92.073602},
  journal = {Phys. Rev. Lett.}
}

@article{emary_Chaos_2003,
  title = {Chaos and the Quantum Phase Transition in the {{Dicke}} Model},
  author = {Emary, Clive and Brandes, Tobias},
  year = 2003,
  month = jun,
  volume = {67},
  number = {6},
  pages = {066203},
  doi = {10.1103/PhysRevE.67.066203},
  journal = {Phys. Rev. E}
}

@article{shapiro_Universal_2020,
  title = {Universal Fluctuations and Squeezing in a Generalized {{Dicke}} Model near the Superradiant Phase Transition},
  author = {Shapiro, D. S. and Pogosov, W. V. and Lozovik, {\relax Yu}. E.},
  year = 2020,
  month = aug,
  volume = {102},
  number = {2},
  pages = {023703},
  doi = {10.1103/PhysRevA.102.023703},
  journal = {Phys. Rev. A}
}

@article{hayashida_Perfect_2023,
  title = {Perfect Intrinsic Squeezing at the Superradiant Phase Transition Critical Point},
  author = {Hayashida, Kenji and Makihara, Takuma and Marquez Peraca, Nicolas and Fallas Padilla, Diego and Pu, Han and Kono, Junichiro and Bamba, Motoaki},
  year = 2023,
  month = feb,
  volume = {13},
  number = {1},
  pages = {2526},
  doi = {10.1038/s41598-023-29202-x},
  journal = {Sci Rep}
}

@article{gleyzes_Quantum_2007,
  title = {Quantum Jumps of Light Recording the Birth and Death of a Photon in a Cavity},
  author = {Gleyzes, S{\'e}bastien and Kuhr, Stefan and Guerlin, Christine and Bernu, Julien and Del{\'e}glise, Samuel and Busk Hoff, Ulrich and Brune, Michel and Raimond, Jean-Michel and Haroche, Serge},
  year = 2007,
  month = mar,
  journal = {Nature},
  volume = {446},
  number = {7133},
  pages = {297--300},
  doi = {10.1038/nature05589}
}

@article{guerlin_Progressive_2007,
  title = {Progressive Field-State Collapse and Quantum Non-Demolition Photon Counting},
  author = {Guerlin, Christine and Bernu, Julien and Del{\'e}glise, Samuel and Sayrin, Cl{\'e}ment and Gleyzes, S{\'e}bastien and Kuhr, Stefan and Brune, Michel and Raimond, Jean-Michel and Haroche, Serge},
  year = 2007,
  month = aug,
  journal = {Nature},
  volume = {448},
  number = {7156},
  pages = {889--893},
  doi = {10.1038/nature06057}
}

@article{saito_Measurementinduced_2003,
  title = {Measurement-Induced Spin Squeezing in a Cavity},
  author = {Saito, Hiroki and Ueda, Masahito},
  year = 2003,
  month = oct,
  volume = {68},
  number = {4},
  pages = {043820},
  doi = {10.1103/PhysRevA.68.043820},
  journal = {Phys. Rev. A}
}

@article{sørensen_Measurement_2003,
  title = {Measurement {{Induced Entanglement}} and {{Quantum Computation}} with {{Atoms}} in {{Optical Cavities}}},
  author = {S{\o}rensen, Anders S. and M{\o}lmer, Klaus},
  year = 2003,
  month = aug,
  volume = {91},
  number = {9},
  pages = {097905},
  doi = {10.1103/PhysRevLett.91.097905},
  journal = {Phys. Rev. Lett.}
}

@article{brif_Phasespace_1999,
  title = {Phase-Space Formulation of Quantum Mechanics and Quantum-State Reconstruction for Physical Systems with {{Lie-group}} Symmetries},
  author = {Brif, C. and Mann, A.},
  year = 1999,
  month = feb,
  volume = {59},
  number = {2},
  pages = {971--987},
  doi = {10.1103/PhysRevA.59.971},
  journal = {Phys. Rev. A}
}

@article{davis_Wigner_2021,
  title = {Wigner Negativity in Spin-$j$ Systems},
  author = {Davis, Jack and Kumari, Meenu and Mann, Robert B. and Ghose, Shohini},
  year = 2021,
  month = aug,
  volume = {3},
  number = {3},
  pages = {033134},
  doi = {10.1103/PhysRevResearch.3.033134},
  journal = {Phys. Rev. Res.}
}

@article{hong-yi_Density_1992,
  title = {Density {{Matrix}} and {{Squeezed State}} of the {{Two Coupled Harmonic Oscillators}}},
  author = {{Hong-Yi}, Fan},
  year = 1992,
  month = jul,
  volume = {19},
  number = {6},
  pages = {443},
  doi = {10.1209/0295-5075/19/6/001},
  journal = {EPL}
}

@article{zhou_Quantum_2020,
  title = {Quantum Entanglement Maintained by Virtual Excitations in an Ultrastrongly-Coupled-Oscillator System},
  author = {Zhou, Jian-Yong and Zhou, Yue-Hui and Yin, Xian-Li and Huang, Jin-Feng and Liao, Jie-Qiao},
  year = 2020,
  month = jul,
  volume = {10},
  number = {1},
  pages = {12557},
  doi = {10.1038/s41598-020-68309-3},
  journal = {Sci Rep}
}

@article{mirkhalaf_Frequency_2025,
  title = {Frequency Shifts Heralding Ground State Squeezing and Entanglement of Two Coupled Harmonic Oscillators},
  author = {Mirkhalaf, Safoura and Ritsch, Helmut and Gietka, Karol},
  journal = {\href{https://doi.org/10.48550/arXiv.2511.03687}{arXiv:2511.03687v2}},
  year = 2025,
}

@article{baumann_Dicke_2010,
  title = {Dicke Quantum Phase Transition with a Superfluid Gas in an Optical Cavity},
  author = {Baumann, Kristian and Guerlin, Christine and Brennecke, Ferdinand and Esslinger, Tilman},
  year = 2010,
  month = apr,
  journal = {Nature},
  volume = {464},
  number = {7293},
  pages = {1301--1306},
  doi = {10.1038/nature09009}
}

@article{mivehvar_Cavity_2021,
  title = {Cavity {{QED}} with Quantum Gases: New Paradigms in Many-Body Physics},
  author = {Mivehvar, Farokh and Piazza, Francesco and Donner, Tobias and Ritsch, Helmut},
  year = 2021,
  month = jan,
  journal = {Advances in Physics},
  volume = {70},
  number = {1},
  pages = {1--153},
  doi = {10.1080/00018732.2021.1969727}
}

@article{kirton_Introduction_2019,
  title = {Introduction to the {{Dicke Model}}: {{From Equilibrium}} to {{Nonequilibrium}}, and {{Vice Versa}}},
  author = {Kirton, Peter and Roses, Mor M. and Keeling, Jonathan and Dalla Torre, Emanuele G.},
  year = 2019,
  journal = {Advanced Quantum Technologies},
  volume = {2},
  number = {1-2},
  pages = {1800043},
  doi = {10.1002/qute.201800043}
}
\end{document}